\documentclass[twocolumn, tighten]{aastex62}

\usepackage{graphicx}	
\usepackage{amsmath}	
\usepackage{amssymb}	
\usepackage[referable]{threeparttablex}

\submitjournal{ApJ}

\shorttitle{Extended Radio AGN in ORELSE}
\shortauthors{Shen et al.}

\begin{document}

\title{EXTENDED RADIO AGN AT z $\sim$ 1 IN THE ORELSE SURVEY: THE CONFINING EFFECT OF DENSE ENVIRONMENTS}

\correspondingauthor{Lu Shen, Guilin Liu, Wenjuan Fang,\\ Hongyan Zhou}
\email{lushen@ustc.edu.cn, glliu@ustc.edu.cn, wjfang@ustc.edu.cn, \\mtzhou@ustc.edu.cn}

\author[0000-0001-9495-7759]{Lu Shen}
\affil{CAS Key Laboratory for Research in Galaxies and Cosmology, Department of Astronomy, University of Science and Technology of China, Hefei 230026, China}
\affil{School of Astronomy and Space Sciences, University of Science and Technology of China, Hefei, 230026, China}

\author{Guilin Liu}
\affil{CAS Key Laboratory for Research in Galaxies and Cosmology, Department of Astronomy, University of Science and Technology of China, Hefei 230026, China}
\affil{School of Astronomy and Space Sciences, University of Science and Technology of China, Hefei, 230026, China}

\author{Meng-Fei Zhang}
\affil{School of Astronomy and Space Science, Nanjing University, Nanjing 210023, China}
\affil{Key Laboratory of Modern Astronomy and Astrophysics, Nanjing University, Nanjing 210023, China}

\author{Brian C. Lemaux}
\affil{Physics Department, University of California, Davis, One Shields Avenue, Davis, CA 95616, USA}

\author{Lori M. Lubin} 
\affil{Physics Department, University of California, Davis, One Shields Avenue, Davis, CA 95616, USA}
\affil{Visiting Scientist, The Observatories, The Carnegie Institution for Science, 813 Santa Barbara St., Pasadena, CA 91101, USA}

\author{Debora Pelliccia} 
\affil{Physics Department, University of California, Davis, One Shields Avenue, Davis, CA 95616, USA}

\author{Emily Moravec}
\affil{Astronomical Institute of the Czech Academy of Sciences, B\v ocn\'i II 1401/1A, 14000 Praha 4, Czech Republic}

\author{Emmet Golden-Marx}
\affil{Department of Astronomy and The Institute for Astrophysical Research, Boston University, 725 Commonwealth Avenue, Boston, MA
02215, USA}

\author{Hongyan Zhou}
\affil{CAS Key Laboratory for Research in Galaxies and Cosmology, Department of Astronomy, University of Science and Technology of China, Hefei 230026, China}
\affil{School of Astronomy and Space Sciences, University of Science and Technology of China, Hefei, 230026, China}

\author{Wenjuan Fang}
\affil{CAS Key Laboratory for Research in Galaxies and Cosmology, Department of Astronomy, University of Science and Technology of China, Hefei 230026, China}
\affil{School of Astronomy and Space Sciences, University of Science and Technology of China, Hefei, 230026, China}

\author{Adam Tomczak} 
\affil{Physics Department, University of California, Davis, One Shields Avenue, Davis, CA 95616, USA}

\author{John McKean}
\affil{Kapteyn Astronomical Institute, University of Groningen, Groningen, the Netherlands}

\author{Neal A. Miller}
\affil{Stevenson University, Department of Mathematics and Physics, 1525 Greenspring Valley Road, Stevenson, MD, 21153, USA}

\author{Christopher D. Fassnacht}
\affil{Physics Department, University of California, Davis, One Shields Avenue, Davis, CA 95616, USA}

\author{Po-Feng Wu}
\affil{National Astronomical Observatory of Japan, Osawa 2-21-1, Mitaka, Tokyo 181-8588, Japan}

\author{Dale Kocevski}
\affil{Colby College, 4000 Mayflower Hill, Waterville, Maine 04901, USA}

\author{Roy Gal}
\affil{University of Hawai'i, Institute for Astronomy, 2680 Woodlawn Drive, Honolulu, HI 96822, USA}

\author{Denise Hung} 
\affil{University of Hawai'i, Institute for Astronomy, 2680 Woodlawn Drive, Honolulu, HI 96822, USA}

\author{Gordon Squires}
\affil{Spitzer Science centre, California Institute of Technology, M/S 220-6, 1200 E. California Blvd., Pasadena, CA, 91125, USA}

\begin{abstract}
Recent hydrodynamic simulations and observations of radio jets have shown that the surrounding environment has a large effect on their resulting morphology. To investigate this we use a sample of 50 Extended Radio Active Galactic Nuclei (ERAGN) detected in the Observations of Redshift Evolution in Large Scale Environments (ORELSE) survey. These sources are all successfully cross-identified to galaxies within a redshift range of $0.55 \leq z \leq 1.35$, either through spectroscopic redshifts or accurate photometric redshifts. 
We find that ERAGN are more compact in high-density environments than those in low-density environments at a significance level of 4.5$\sigma$. Among a series of internal properties under our scrutiny, only the radio power demonstrates a positive correlation with their spatial extent. After removing the possible radio power effect, the difference of size in low- and high-density environments persists. In the global environment analyses, the majority (86\%) of high-density ERAGN reside in the cluster/group environment. In addition, ERAGN in the cluster/group central regions are preferentially compact with a small scatter in size, compared to those in the cluster/group intermediate regions and fields. 
In conclusion, our data appear to support the interpretation that the dense intracluster gas in the central regions of galaxy clusters plays a major role in confining the spatial extent of radio jets. 

\end{abstract}


\section{Introduction} \label{sec:intro}

The relation between Radio Active Galactic Nuclei (RAGN) and their large-scale environments has been explored in numerical simulations and observational programs. RAGN are preferentially found in dense cluster environments at low redshift (e.g., \citealp{Miller2002, Best2004, Argudo2016}) and such a preference appears to persist up to z $\sim$ 2 (e.g., \citealp{Hatch2014, Malavasi2015, Magliocchetti2016, Shen2017, Mo2018}). A relationship between radio luminosity and environmental richness has been revealed by a number of works (e.g. \citealp{Best2004, Ineson2013, Ineson2015, Ching2017, Croston2019}). In general, richer environments appear to host more luminous radio galaxies, with such a relationship possibly linked both to AGN accretion mode \citep{Ineson2015} and to radio morphology \citep{Croston2019}. 

In addition to the linkage between radio luminosity and the environment, the interaction between the jet powered by the RAGN and the intracluster medium (ICM) can potentially provide insights into the RAGN - environment complexity. 
There are several channels that may enable RAGN to efficiently interact with the ICM, including the displacement of gas, shocks, or the transportation of low entropy gas and heavy elements outward from the cluster cores (see \citealp{Fabian2012} for review).  
Meanwhile, it has been argued that group or cluster-like external pressures are required in order to provide a medium to confine the expanding radio-lobe plasma, at least for RAGN observed in the local universe (e.g., \citealp{McNamara2000, Fabian2000, Johnstone2002, Fabian2005, Forman2007}). 
In addition, hydrodynamic simulations of radio jets have shown that, for a given jet power, the cluster or group environment in which the radio jet is propagating affects the resulting radio morphology, lobe dynamics and other observable properties \citep{Yates2018}. 
Recently, \citet{Moravec2019, Moravec2020} found evidence that the size of the jets increases with the clustocentric radius, after investigating $\sim$50 extended RAGN located in massive galaxy clusters detected by infrared-selected galaxy overdensities at z $\sim$ 1. 
They argued that more compact jets occur near the center where the ICM pressure is the highest and confines the jets. 
However, while they identified low redshift interlopers through a color-color analysis, due to the lack of redshifts of individual galaxies, they assumed that each radio source was at the redshift of the cluster. This assumption introduces an uncertainty on the real membership of their extended RAGN, and thus potentially complicates the interpretation of their results.

While attempts to understand the environmental effects on the size of extended RAGN (ERAGN) have been made, very few observational investigations across a wide range of environments (i.e., from cluster center to the field) have been conducted outside of the local universe where the relaxed fraction of clusters starts to decrease and ICM profiles become more complex.
Thanks to its well-defined wide dynamic range of environments and its spectroscopic and photometric redshift measurements with high-degree of accuracy, the Observation of Redshift Evolution in Large Scale Environments (ORELSE; \citealp{Lubin2009}) survey facilitates such an investigation. Hence, we study a sample of 50 ERAGN over a wide range of local and global environments at intermediate redshifts (0.55 $\leq z \leq$ 1.3) across 12 ORELSE fields. In this paper, we describe the available ORELSE catalogs, the radio observations, sample selection, and the measurement of radio size in Section \ref{sec:data}. In Section \ref{sec:results}, we show our results on the environmental effect of the size of ERAGN and analyses of its potential causes. In Section \ref{sec:simulation}, we conduct a simulation based on our observations. We conclude with a summary in Section \ref{sec:summary}. Throughout this paper, we adopt the AB system for all magnitudes \citep{Oke1983, Fukugita1996}, a concordance $\Lambda$CDM cosmology with $\mathrm{H_0 = 70\ km\ s^{-1} Mpc^{-1}}$, $\Omega_\Lambda = 0.73$, and $\Omega_{\mathrm{M}} = 0.27$, and a Chabrier stellar initial mass function (IMF; \citealp{Chabrier2003}).

\section{DATA AND SAMPLE SELECTION}
\label{sec:data}

\begin{deluxetable}{lccccc}
\tablecaption{Depth of Imaging Data \label{tab:fields}} 
\tablehead{\colhead{Field} & \colhead{R.A.} & \colhead{Decl.} & \colhead{$<z_{\text{spec}}>$} & \colhead{Depth} & \colhead{Num of} \vspace{-2mm}\\
& & & & \colhead{($\mu$Jy)}  &\colhead{ERAGN} \vspace{-2mm}\\
 \colhead{(1)} & \colhead{(2)} &  \colhead{(3)} &  \colhead{(4)}  & \colhead{(5)}  &\colhead{(6)}}
\startdata
    	SG0023 & 00:24:29 & +04:08:22 & 0.845 & 13.9 & 4  \\
	XLSS005 & 02:27:10 & -04:18:05 & 1.056 & 14.1 & 2+3 \\
	SC0910 & 09:10:45 & +54:22:09 & 1.110 & 14.7 & 1 \\
	RXJ1053 & 10:53:40 & +57:35:18 & 1.140 & 10.2 & 3+1 \\
	Cl1137 & 11:37:33 & +30:00:04 & 0.955 & 7.3 &3 \\ 
	SC1324 &13:24:46 & +30:34:11 & 0.717 & 11.0  &10 \\ 
    	Cl1350 & 13:50:48 & +60:07:07 & 0.804 & 9.8  & 2 \\ 
	Cl1429 & 14:29:06 & +42:41:02 &  0.987 & 17.1  & 3+1 \\
	SC1604 & 16:04:15 & +43:21:37 & 0.898 & 9.3  & 9 \\ 
	RXJ1716 & 17:16:50 & +67:08:30 & 0.813 & 15.2  & 2+2 \\ 
	RXJ1757 & 17:57:20 & +66:31:32 & 0.693 & 10.5  & 1 \\
	RXJ1821 &  18:21:38 & +68:27:52 & 0.818 & 9.3 &4 \\
\enddata
\tablenotetext{}{(1) (2) (3) Field and the coordinates of center of radio imaging. (4) Mean spectroscopic redshift of the main structure in each field. (5) The depth of observations is the source-free RMS in the full image area derived from the task SAD in the AIPS. Due to the variation of RMS across each pointing, this corresponds to $\sim$15 arcmin from the center of that pointing. ERAGN are located on average 6.6$\arcmin$ from the center of fields and in the range of $0.1\arcmin\sim21.6\arcmin$. (6) Numbers of ERAGN automatically-detected in the radio catalogs plus those detected by eye that are matched to optical counterparts. }
\end{deluxetable}

\subsection{The ORELSE Survey and Available Data}
\label{sec:ORELSE}

In this paper, we use the 12 fields from the ORELSE survey that have fully reduced radio, photometric, and spectroscopic catalogues. 
The construction of the existing data and photometric parameters adopted in this paper are described in \citet{Tomczak2017, Tomczak2019}. Spectral Energy Distribution (SED) fitting is performed on deep multi-wavelength imaging to estimate photometric redshift ($z_{phot}$), restframe color, stellar masses as well as other properties of the stellar populations of galaxies. 
The spectroscopic redshifts catalogs are extracted and assessed primarily on observations from the DEep Imaging and Multi-Object Spectrograph (DEIMOS;~\citealp{Faber2003}) equipped on Keck II. Complementary spectroscopy is obtained in the literature \citep{Oke1998, Gal2004, Gioia2004, Tanaka2008}, and the different facilities utilized therein are detailed in \citealp{Lemaux2012, Lemaux2019}.

\subsection{Radio Observations and Sample Selection}
\label{sec:sample} 

All of the 12 fields under consideration here were mapped using the Karl G. Jansky Very Large Array (VLA) at 1.4GHz in its B configuration, where the synthesized beam is about 5'' (FWHM) and the field of view (i.e., the FWHM width of the primary beam) is approximately 31' in diameter.  
Data reduction and source catalogs for the fields (SC1604, SG0023, SC1324, RXJ1757 and RXJ1821) mapped with the traditional VLA are described in \citet{Shen2017, Shen2019}. Those of the remaining seven fields (XLSS005, RXJ1053, RXJ0910, RXJ1716, Cl1137, Cl1429 and Cl1350) mapped with the new JVLA are obtained following the same methodology as in \citet{Shen2020}. 
For those catalogs, we automatically detect radio sources down to a 4$\sigma$ detection flux density limit, where $\sigma$ is the local RMS noise. The relevant information is summarized in Table \ref{tab:fields}. 

When we extract extended radio sources from the constructed catalogs, these sources are required to subtend a major axis of at least 6.5'' (50 kpc at z = 1), a threshold 30\% larger than the FWHM = 5'' resolution of our VLA data. 
We then search for additional radio sources that contain multiple components by eye. The task \textsc{IMFIT} in the NRAO's Astronomical Image Processing System (AIPS) is employed to fit multiple Gaussian components simultaneously to these sources. The total integrated flux density is the combined integrated flux density of the individual components, and the associated error is the square-sum of their errors. The center is either the halfway point between the flux peak of each component for a two component system or the flux peak of the middle component for a three components system. 

To identify the optical counterparts to these extended radio sources, we perform a maximum likelihood ratio (LR) technique following (\citealp{Rumbaugh2012}; Section 3.4). We adopt a search radius of $2\arcsec$ between the overall photometric catalogs and the extended sources to account for astrophysical and astrometric offsets. We note that moving to a $5\arcsec$ search radius did not yield any additional sources. 

Considering our local overdensity footprints (see Section \ref{sec:env}), the completeness/representativeness of our spectroscopic catalogs, and consistencies with previous radio galaxy studies in ORELSE, we require host galaxies to have 0.55 $\leq$ z $\leq$ 1.35, $18.5 \le i'/z \le 24.5$ and $\mathrm{M_* \geq 10^{10}M_\odot}$ (\citealp{Shen2017, Lemaux2019}). As a result, we have a total sample of 50 ERAGN (Table \ref{tab:fields}) of which 24 have spectroscopically-confirmed redshifts (hereafter ERAGN-spec) and the rest have accurate photometric redshifts (hereafter ERAGN-phot). 
We note that the stellar mass cut excludes 4 ERAGN-phot. However, the main conclusions do not change if we instead included all ERAGN. 
From the original design of the ORELSE survey, we have preferentially targeted galaxies that might reside in the clusters/groups for spectroscopic observations. In this study, we use both spectroscopic and photometric galaxies and account for their associated $z_\mathrm{phot}$ to mitigate this selection effect. In addition, the high-priority targets (red sequence cluster members) are always sub-dominant in our observations \citep{Lemaux2019}, and the resultant spectral galaxy sample is mostly representative of the underlying galaxy population at these redshifts \citep{Shen2017, Lemaux2019}. 

\begin{figure*}
	\centering
	\includegraphics[width=\textwidth]{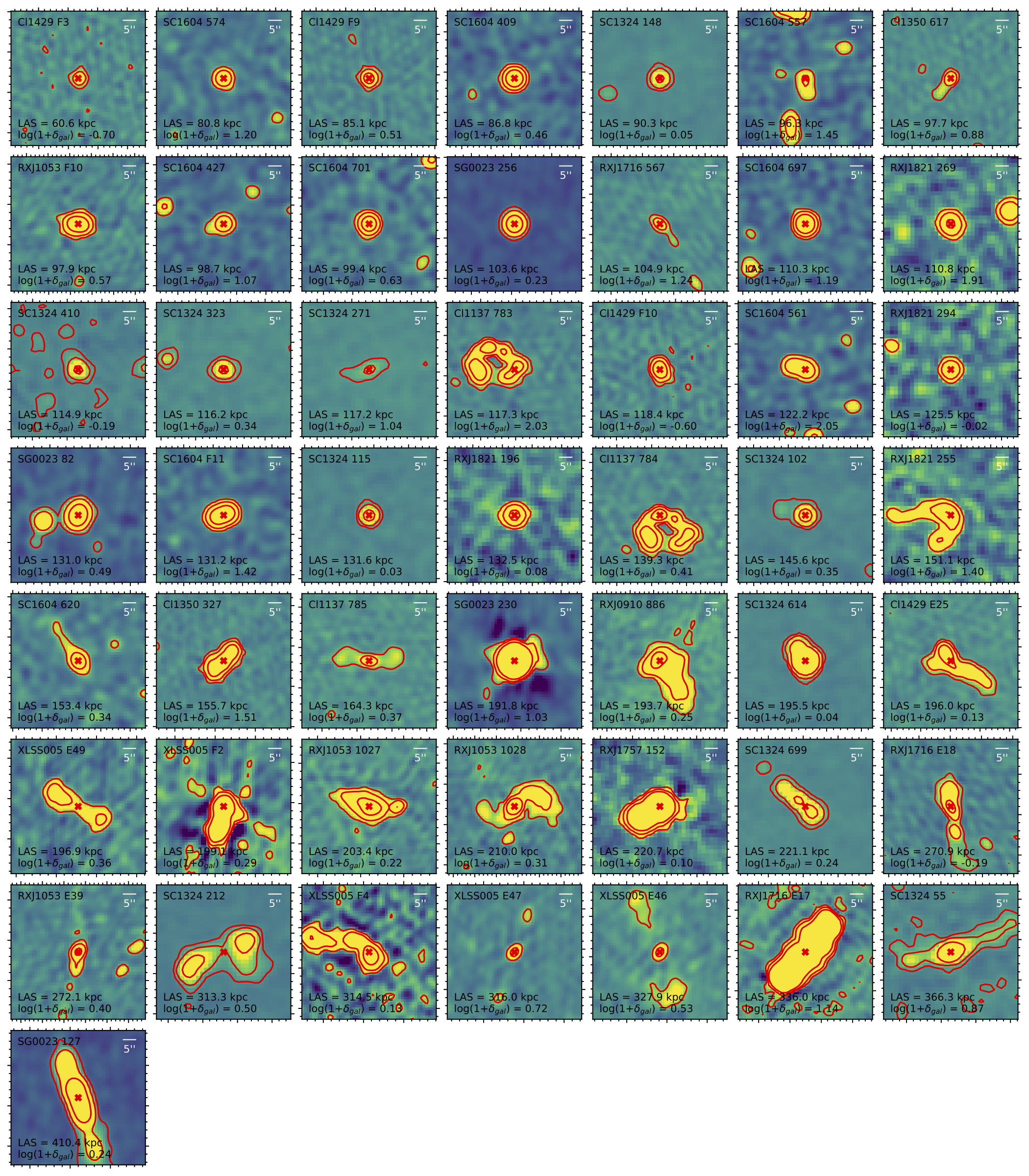}
    \caption{Radio cutouts of ERAGN ranked by their radio size in ascending order. Images are 50'' $\times$ 50'' (400 kpc $\times$ 400 kpc at $z$ = 1) indicated by the scale bar in the upper right hand corner, and are centered on the centers of radio sources that are marked by red crosses. North is up and east is to the left. The contours levels are 4, 16 and 64 $\sigma$. The labels are field name plus radio id in the upper left, and LAS and local overdensity values in the lower left. }
    \label{fig:radio}
\end{figure*}

We expect high completeness and purity of the ERAGN sample. The fluxes of the ERAGN are all above the 80\% completeness level of the overall radio sources in the photometric footprint of the 12 fields, and the fluxes of those ERAGN detected by eye are above the 95\% completeness level. Thus, we are not biased by the different depth of radio images and local RMS variance. As for the purity, all ERAGN are detected $\geq 33\sigma$ significant, and $\geq 90\sigma$ for those ERAGN detected by eye, which are much higher than the 4$\sigma$ detection limit of the automatically-detected radio sources. They are also associated with an optical counterpart. Thus, it is unlikely these radio detection are spurious. Another possible case which might affect the purity is that two blended sources are detected as a single extended source. Here, we relied mostly on the visual check from the contour plots and do not think any of them could be this case.
We do not find any correlation between size and redshift in the ERAGN sample (see discussion in Section \ref{sec:hosts}), so adopting a physical (i.e. kpc) or angular size cut does not affect our selection and results. 
We show their radio map cutouts of our sample in Figure \ref{fig:radio}, and overlay radio contours on their optical identification images in Figure \ref{fig:optical}. 
The properties of 50 ERAGN are presented in Appendix \ref{sec:catalog} (Table \ref{tab:catalog}). 

Radio power is calculated in the same way as used in previous radio galaxy studies in ORELSE:
$$L_{1.4GHz} = 4\pi D_L^2 S_{1.4GHz} (1 + z)^{(\alpha - 1)}$$
where $D_L$ is the luminosity distance at the redshift, $S_{1.4GHz}$ is the total integrated radio flux, and $(1 + z)^{(\alpha - 1)}$ includes both the distance dimming and K-correction. To be consistent to previous radio galaxy studies in ORELSE, we adopt the same $\alpha = 0.7$. Typical values of spectral index $\alpha$ for extended radio sources range from 0.5 to 1 \citep{Condon1992, Peterson1997, Lin2007, Miley2008, Chhetri2012}. The associated error on the radio power is calculated from the error of total integrated radio flux. To account for the uncertainty of z$_\text{phot}$ of ERAGN-phot, we add an extra error on the L$_\text{1.4GHz}$ by adopting a Monte-Carlo sampling method based on the z$_\text{phot}$ error (see Appendix \ref{app:eta} for the description of this method). 

\begin{figure*}
	\centering
	\includegraphics[width=\textwidth]{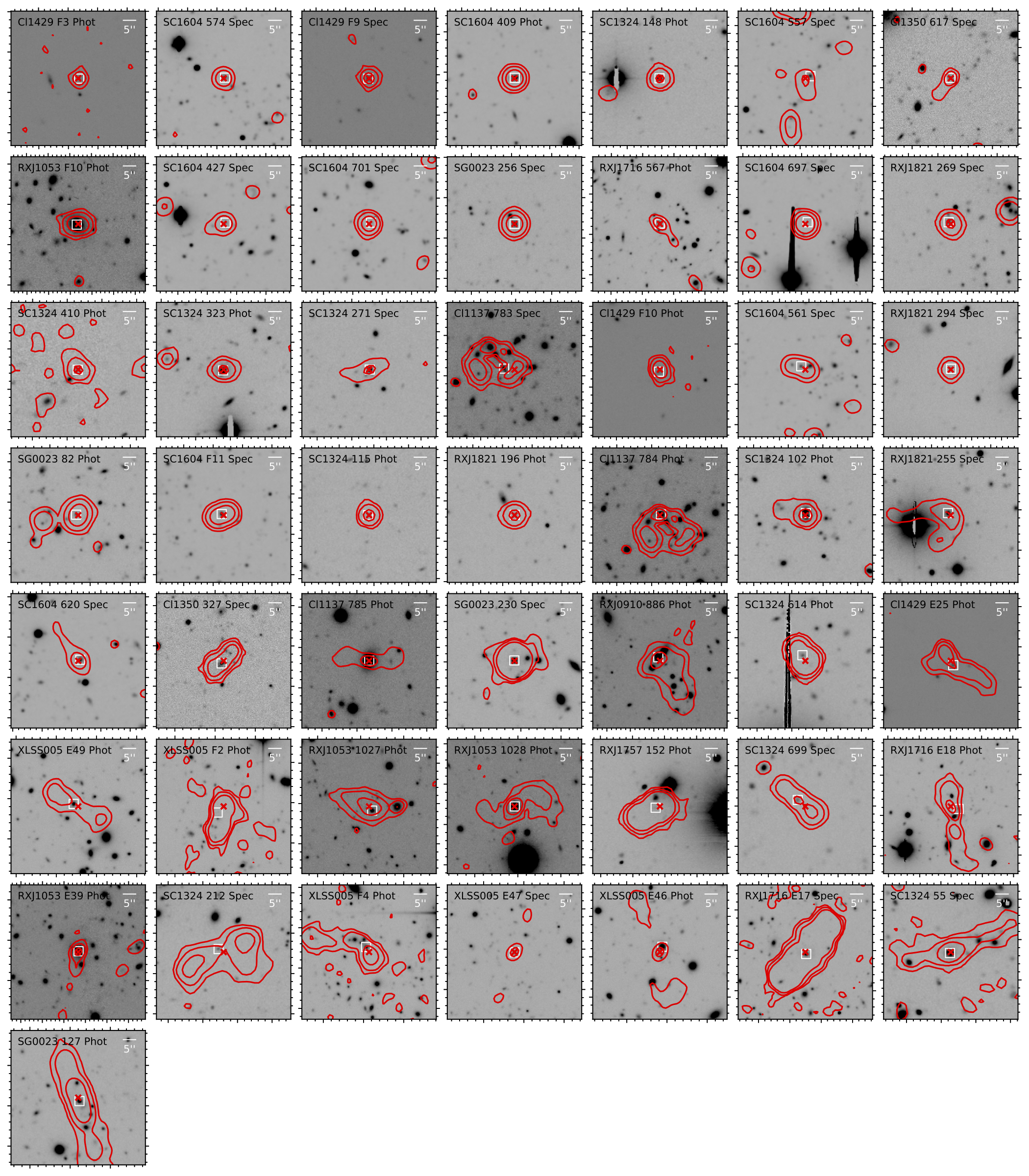}
    \caption{Cutouts of the optical identification images centered on ERAGN ranked by their radio size in ascending order. Images are 50'' $\times$ 50'' (400 kpc $\times$ 400 kpc at $z$ = 1) indicated by the scale bar in the upper right hand corner. The centers of the radio sources are marked by red crosses, and their optical counterparts are marked by white boxes. North is up and east is to the left. The radio contours levels are 4, 16 and 64 $\sigma$. The labels in the upper left are field name, radio id and matched to spectroscopically-confirmed or photometric galaxies.}
    \label{fig:optical}
\end{figure*}

\subsection{Extent of Radio Sources}
\label{sec:LAS}

We adopt the largest angular size (LAS) as the primary measurement of the extent of radio sources, following \citet{Moravec2019}. It is defined as the largest angular distance between a pair of points that belongs to this radio source. 
For the majority of our sample, we measure the LAS from the 4$\sigma$ radio contours, the same as our radio detection threshold \citep{Shen2017, Shen2019}. In case of blending of radio sources, we adopt a higher threshold so that deblending barely happens, which is 8$\sigma$ for SG0023+82 and XLSS005+F4, and 32$\sigma$ for Cl1137+784 and Cl1137+783. In Figure \ref{fig:size_env} and \ref{fig:global}, these four ERAGN are marked by black open diamonds. Note that they are not outliers in any other property analyzed in this paper, and excluding these four sources would not affect any of our results. The uncertainty in a LAS is defined as half of the beam size base on the Nyquist limit of our radio observations. For ERAGN-phot, we add an extra error on the LAS based on the 16/84 percentile values of the mock LAS distributions calculated from the Monte-Carlo z$_\text{phot}$ samplings (see Appendix \ref{app:eta}). 

Considering that the LAS is sensitive to the dynamic variety of radio morphology, we also calculate the area of the radio sources to enhance the robustness of our analysis. This is done by counting the number of pixels enclosed by the same contours as used in the LAS measurements, including the fractional pixels. The uncertainty in an area is defined as the square root of the number of pixels based on the Poisson uncertainty. For ERAGN-phot, we add an extra error on the area (in physical units) based on the 16/84 percentile values of the mock area distributions calculated from the Monte-Carlo z$_\text{phot}$ samplings (see Appendix \ref{app:eta}).

\subsection{Environmental Measurements} 
\label{sec:env}

The environment around a galaxy has two levels: in terms of the cosmic time, the `local' or `global' environment possesses the current or time-averaged density field to which a galaxy has been exposed, respectively. Two environment measurements have been introduced in ORELSE to quantify this: the local environment, defined as $\mathrm{log(1 + \delta_{gal})}$, is obtained using a Voronoi Monte-Carlo (VMC) algorithm which is described in full detail in \citet{Lemaux2017, Tomczak2017, Hung2020}; while the global environment is defined as $\eta = \mathrm{R_{proj}/R_{200}} \times |\Delta v|/\sigma_v$ following the method described in \citet{Shen2019}. In this paper, we adopt $\eta \leq 0.1$ as cluster/group center region, $0.1 < \eta \leq 2$ as intermediate region, and $\eta > 2 $ as field region based on calibrations from N-body simulations \citep{Noble2013, Noble2016}.
The final cluster/group catalog includes the pre-existing spectroscopically confirmed clusters/groups \citep{Gal2008} and overdensity candidate regions \citep{Hung2020}. The details of the uncertainties in the global environment measurement, due to the differences in the cluster/group catalogs, the uncertainties of cluster/group properties and photometric redshifts, are fully discussed in Appendix \ref{app:eta}. 
To account for the uncertainties of z$_\text{phot}$ in local environment measurement, we add an error on the log($1+\delta_\mathrm{gal}$) of the ERAGN-phot based on the 16/84 percentile values of their distributions calculated from the 100 Monte-Carlo z$_\text{phot}$ samplings (see Appendix \ref{app:eta}). Since the uncertainties of z$_\mathrm{spec}$ are smaller than the redshift difference of VMC maps, we do not include errors on $\mathrm{log(1 + \delta_{gal})}$ for the ERAGN-spec. 

In general, regions with $\mathrm{log(1 + \delta_{gal})} < 0.5$ correspond to field-like environments, and regions with $0.5 \leq \mathrm{log(1 + \delta_{gal})} < 1$ are group-like environments, while regions with $\mathrm{log(1 + \delta_{gal})} \ge 1$ are cluster-like environments (see \citealp{Tomczak2017, Tomczak2019}). However, there is certainly not a direct one-to-one correlation between $\delta_{gal}$ and the structure in which a galaxy resides. Thus, to study them together, we analyze both the local, short-term environmental effects and global, long-term environmental effects. 

\section{The Environmental Effect on the Radio Size}
\label{sec:results}

\begin{figure*}
	\includegraphics[width=\textwidth]{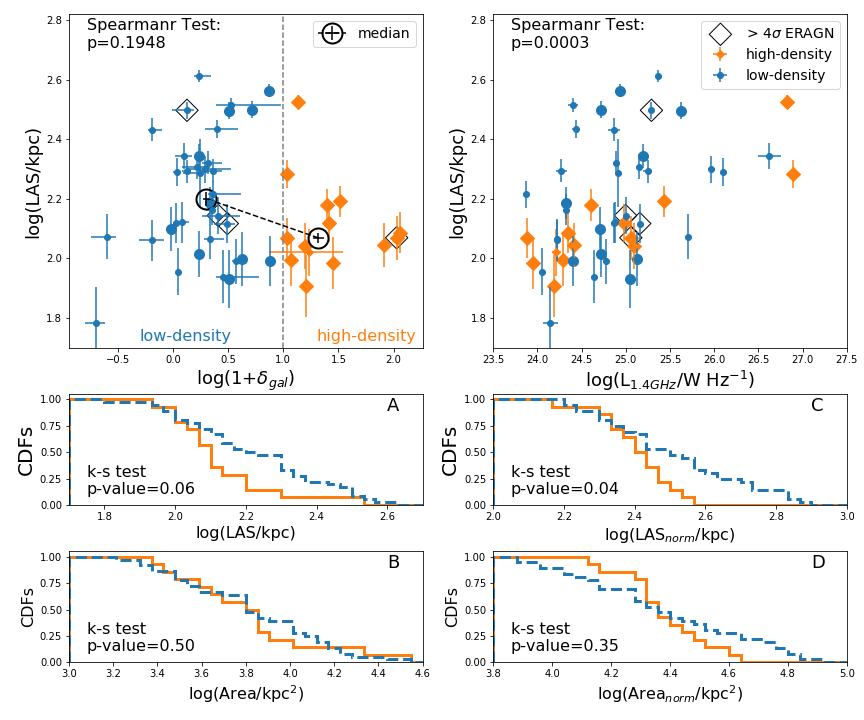}
    \caption{\textit{Left tops:} The LAS versus the local overdensity for low- and high-density ERAGN samples in blue dots and orange diamonds, respectively.  Larger markers are ERAGN having spectroscopically-confirmed redshifts. Four ERAGN using $>4\sigma$ contours to measure their LAS, due to blending (see Section \ref{sec:LAS}), are marked by black diamonds in the top two panels. The black dots are the median of the log(LAS) and log(1 + $\delta_{\text{gal}}$) of low- and high-density ERAGN with error on median. The vertical dashed line shows the separation of low- and high-density at log(1 + $\delta_{\text{gal}}) = 1$.  The median of log(LAS) of the high-density sample is lower than that of the low-density at 4.5$\sigma$ significance level.  \textit{Right top:} The LAS versus the radio power for the low- and high-density ERAGN samples. The significance of the Spearman test (p$_{\mathrm{Spearmanr}}$) for the full ERAGN sample is shown in the top two panels, which suggests a correlation between the radio power and size, but no correlation between the local overdensity and size. \textit{Bottom:}  CDFs of log(LAS) (\textit{A}), log(Area) (\textit{B}), radio power normalized LAS (LAS$_{\text{norm}}$; \textit{C}) and area (Area$_{\text{norm}}$; \textit{D}) for the low- and high-density ERAGN, in blue dashed and orange lines, respectively. Both p-values for the LAS and LAS$_{\text{norm}}$ suggest that the size distributions of low- and high-density ERAGN are drawn from different underlying populations, while it is inconclusive in the area comparisons. }
    \label{fig:size_env}
\end{figure*}

In the left top panel of Figure \ref{fig:size_env}, we plot the radio size against the local overdensity for the full ERAGN sample. A lack of large radio sources in dense regions is evident. 
To quantify this lack, we divide the full sample into low- and high-density subsamples at log(1 + $\delta_{\text{gal}}) = 1$. The median of these two sub-samples are depicted by black dots with errorbars\footnote{Errors on median are calculated as $\sigma = \sigma_{\text{NMAD}}$/$\sqrt{n-1}$, where $\sigma_{\text{NMAD}}$ is the normalized median of the absolute deviations \citep{Hoaglin1983} and n is the number of the sample. }. 
The log(LAS) median of the high-density bin is 2.07 $\pm$ 0.02, in contrast to 2.20 $\pm$ 0.02 for the low-density bin, leading to a discrepancy at a 4.5$\sigma$ significance level. 
There are 2 out of 14 (14\%) high-density ERAGN with a LAS larger than 160 kpc, while 17 out of 36 low-density ERAGN (47\%) are larger than this threshold. 
We quantify this lack of large ERAGN in the dense regions using the following Monte-Carlo approach. In each realization, we randomly sample the same number of high-density ERAGN from the LAS distribution of the low-density sample. Only 1\% of the 1000 realizations have less than or equal to two ERAGN with a LAS larger than this threshold. Therefore, it is unlikely that the high-density ERAGN are just a random sub-sampling of those ERAGN in the low-density sample.

The cumulative distribution functions (CDFs) of log(LAS) for the low- and high-density ERAGN are shown in the left bottom panel A of Figure \ref{fig:size_env}. We perform a Kolmogorov-Smirnov (K-S) test to assess the similarity of two distributions and adopt p $<$ 0.1 as the suggestive threshold for judging whether the samples are drawn from different distributions. The small value of p = 0.06 found in the LAS comparison again suggests that the size distribution of low- and high-density ERAGN are likely drawn from different distributions. Because this p value does not definitively reject the null hypothesis, we attempt further tests below to explore the potential differences. 

We investigate the possible correlation between the local overdensity and LAS based on a non-parametric Spearman rank correlation test. 
We regard a returned p $<$ 0.1 as the suggestive criterion for the existence of a correlation and p $<$ 0.003 as 3$\sigma$ significant threshold. We run the Spearman rank correlation test using the entire ERAGN sample. The large p value, as shown in the top corner of the plot, suggests no correlation between these two properties. 
However, the difference between the size of ERAGN in high- and low-density is significant. We further perform several diagnostic tests to eliminate uncertainties associated with the LAS measurement, binning, and projection effects. 

To account for the uncertainty associated with individual LAS values, we adopt a Monte-Carlo approach, where 10000 realizations of the median are computed after randomly perturbing each LAS value as per their errors. The median LAS of high- and low-density ERAGN are $121.0 \pm 7.5$ and $164.2\pm 8.0$ kpc, respectively, where the errors are the 16th/84th percentiles. Hence, in terms of medians, the LAS of high-density ERAGN is smaller than that of low-density ERAGN by a factor of 1.4 at a significance level of 4$\sigma$. 

To test the robustness of the above results due to different binning, we vary the log(1$+\delta_{\text{gal}}$) threshold successively from 0.1 to 1.5 with a step of 0.1, and perform a K-S test in each step. The p value is found to be consistently below 0.1 when log(1 + $\delta_{\text{gal}}$) reaches $\sim$ 0.4, indicating that the difference of LAS distributions is robust against binning variation. 
We also noticed that one of the high-density ERAGN might change to the low-density sub-sample due to its large error on log($1+\delta_\mathrm{gal}$) (see Figure \ref{fig:size_env}). However, none of our results would be meaningfully affected if this ERAGN was classified in the low-density sub-sample. 

Lastly, we test the potential projection effects using a bootstrap approach. To do this, we assume that the low-density ERAGN subsample represent the overall distribution of radio sources with random viewing angles. We randomly resample the 14 LAS values from the low-density ERAGN subsample and calculate the median and spread\footnote{The difference between 16th and 84th percentile} for 10000 realizations. As a result, our bootstrapping leads to a distribution of log(LAS) with a median of 2.18, larger than the measured value of the actual high-density ERAGN in 9570 out of the 10000 realizations. The spread of log(LAS) is 0.42, larger than the measured value of high-density ERAGN in 9930 realizations. 

As mentioned in Section \ref{sec:LAS}, we characterize the size of the radio sources using their area as a complementary tracer. The CDFs of log(area) for the low- and high-density ERAGN are shown in the left bottom panel B of Figure \ref{fig:size_env}. The K-S test result from the area comparison is inconclusive. We perform the same Monte-Carlo sampling as run in the LAS analyses. We find only 4\% of the realizations where there were less than or equal to two ERAGN with an emitting region larger than 10$^4$ kpc$^2$, to compare to the two observed high-density ERAGN that are larger than this threshold. Together with the differences shown in the LAS comparison, these results hint at the emitting regions of high-density ERAGN being more isotropic than those of the low-density ERAGN. 

These tests confirm the existence of the size difference between high- and low-density ERAGN, a phenomenon likely a consequence of the host galaxies internal properties and intracluster environment. Previous works have shown that the length of radio jets depend on the jet power, the local density profile and the time that has elapsed since the jets started (e.g., \citealp{Falle1991, Hardcastle2018, Yates2018}), and that massive galaxies reside in the inner portion of clusters and tend to host compact jets \citep{Moravec2020}. Our discussion on the potential effects due to hosts, radio power and the large-scale environment are presented below.  

\subsection{Effect of Hosts}
\label{sec:hosts}

We examine the dependence of LAS of ERAGN on the stellar mass, quiescent state and redshift of their galaxy hosts. 
Whether a host belongs to the quiescent or star-forming regime is determined by its location on the rest-frame $\mathrm{M_{NUV}} - \mathrm{M_r}$ versus $\mathrm{M_r} - \mathrm{M_J}$ color-color diagram. We adopt separations from \citet{Lemaux2014} \footnote{Galaxies at $z\leq1$ with $\mathrm{M_{NUV} - M_r > 2.8(M_r - M_J)+1.51}$ and $\mathrm{M_{NUV}  - M_r > 3.75}$ and galaxies at $z>1$ with $\mathrm{M_{NUV} - M_{r} > 2.8(M_r - M_J)+1.36}$ and  $\mathrm{M_{NUV} - M_r >3.6}$ are considered quiescent. }
and calculated a color offset as the perpendicular offset from the quiescent/star-forming separation line, with positiveness representing the quiescent regime (\citealp{Shen2017}).

We run Spearman rank correlation tests on the relationship between the stellar mass/color offset/redshift and LAS using the entire ERAGN sample.
All p values turn out to be $>$0.1 (p = 0.73/0.72/0.48 for stellar mass/color offset/redshift), which suggests no correlation between the size and the host internal properties. In conclusion, the galaxy host internal properties neither have significant effects on the radio size nor introduce significant bias to our main result. 

\subsection{Effect of Radio Power}
\label{sec:radiopower}

In the top right panel of Figure \ref{fig:size_env}, we plot the LAS with the radio power (L$_\text{{1.4GHz}}$). Overall, the size increases with increasing radio power for the entire sample. The small p value obtained in Spearman rank test (p = 0.0003) implies the existence of a correlation between the two variables at $> 3\sigma$ level; while $\rho$ = 0.5 suggests a positive correlation between the two variables. 
Consequently, the radio power likely acts as a driver of the size of ERAGN. 
This is intuitive as the further a jet is driven, the more energetic the outburst is likely to be. This conclusion is in line with simulations reporting that the length of jets depends on the jet power and is positively correlated with radio luminosity (e.g., \citealp{Falle1991, Hardcastle2018, Yates2018}). 

To alleviate this effect on the size difference between high- and low-density ERAGN, we normalize the LAS and area to correct for their radio power following the method in \citet{Moravec2019} that 
$$\mathrm{LAS_{norm} = LAS(\frac{L_0}{L_{1.4GHz}})^{1/5}},$$
$$\mathrm{Area_{norm} = Area(\frac{L_0}{L_{1.4GHz}})^{2/5}}, $$where L$_0$= 2 $\times$ 10$^{26}$ W Hz$^{-1}$ and L$_\mathrm{{1.4GHz}}$ is the radio power at 1.4GHz. 
The errors associated with each quantity are calculated by standard propagation of the individual LAS/Area and L$_\mathrm{{1.4GHz}}$ errors. The CDFs of LAS$_\text{{norm}}$ and area$_\text{{norm}}$ for low-/high-density ERAGN are shown in right bottom panel C and D of Figure \ref{fig:size_env}.
These corrections result in a more obvious difference between the high- and low-density ERAGN subsamples, compared to the non-normalized LAS and the area measurement.  Consistently, we find a smaller p = 0.04 returned from the K-S test performed on the two log(LAS$_\text{{norm}}$) distributions. 
The K-S test result is still inconclusive in the area comparison.

\subsection{Effect of Cluster Environments} 
\label{sec:global}

\begin{figure*}
    \centering
    \includegraphics[width=\textwidth]{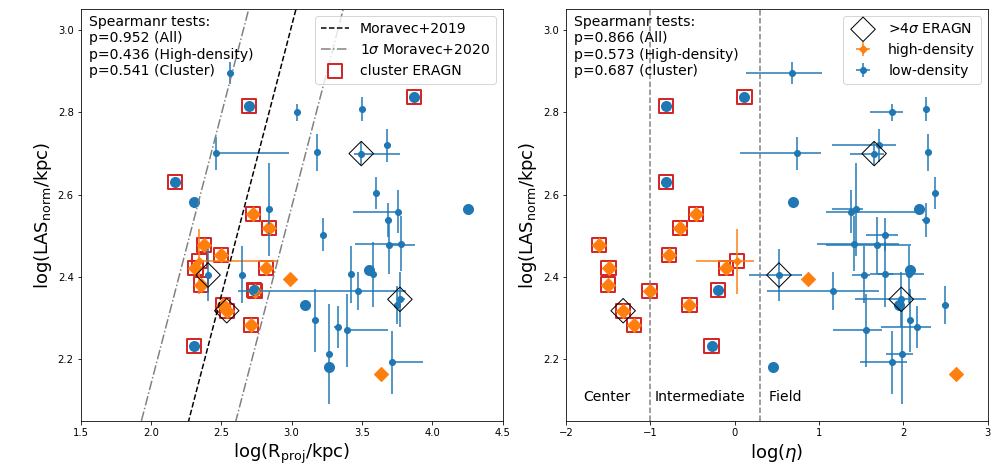}
    \caption{The log(LAS$_{norm}$) versus the distance from the cluster center (\textit{Left}) and global environment $\eta$ (\textit{Right}) for the low-density ERAGN in blue dots and high-density ERAGN in orange diamonds. ERAGN within clusters/groups (cluster-ERAGN, log($\eta$) $\leq$ 2) are marked with red open boxes. The errorbars on R$_\text{proj}$ and log($\eta$) of ERAGN-phot are derived based on the uncertainty of z$_\text{phot}$ (see Appendix \ref{app:eta}). Larger markers are ERAGN having spectroscopically-confirmed redshifts. Four ERAGN using $>4\sigma$ contours to measure their LAS, due to blending (see Section \ref{sec:LAS}), are marked by black open diamonds in both panels. The black dashed line is the relationship from \citet{Moravec2019}, along with 1$\sigma$ (grey dashed lines) derived from the full sample in \citet{Moravec2020}. The cluster/group center, the intermediate and field regions are separated by the vertical dashed lines. The p values of the Spearman rank correlation tests of the full, high-density, cluster-hosted ERAGN samples are shown in the upper left corner of each panel. None of them shows a noticeable correlation. }
\label{fig:global}
\end{figure*}

The ORELSE cluster/group catalogs allow for investigating the large-scale environmental preference of ERAGN through studying their relationship to the parent cluster/group structures. 
Indeed, the vast majority of high-density ERAGN (86\%) reside in the cluster/group environment of which 6 are even located in the cluster/group centers. In contrast, only 14\% of low-density ERAGN reside in the cluster/group environment with none of them in the cluster/group centers. 
In the Figure \ref{fig:global}, we show the LAS$_\text{norm}$ as a function of clustocentric radius ($R_\text{proj}$) and global environment (log($\eta$)) for the low- and high-density ERAGN subsamples, respectively. ERAGN within clusters/groups (cluster-ERAGN, log($\eta$) $\leq$ 2) are marked with red open boxes.

We perform Spearman rank correlation tests on R$_\mathrm{proj}$ and LAS$_\text{norm}$ of the high-density, cluster-hosted, and full ERAGN samples, shown in the upper left corner of Figure \ref{fig:global}. None of them show a significant correlation. 
This result is inconsistent with \citet{Moravec2019} who find a tight LAS vs. clustocentric radius relation at z $\sim$ 1 (black dashed line). 
\citet{Moravec2020} confirmed this trend with a scatter ($\sigma$ = 0.44, shown as the grey dashed lines in Figure \ref{fig:global}) using a sample of 50 ERAGN selected within 1$\arcmin$ in projected clustocentric radius for clusters at z$\sim$ 1 in the Massive and Distant Clusters of WISE Survey  (MaDCoWS; \citealp{Gonzalez2019}). 
In addition, Golden-Marx et al. (\textit{in prep.}) also find little agreement with this trend by analyzing 36 extended bent RAGN that reside in clusters detected at similar redshifts in the high-z Clusters Occupied by Bent Radio AGN survey (COBRA; \citealp{Golden-Marx2019}). 

It is known that the ICM is plausibly more developed in more massive clusters and at lower redshift at fixed mass, and their ICM density profile is more regular \citep{Newman2013, Rumbaugh2018}. As a consequence, the effect of a more developed cluster environment on the size might show a better correlation with the clustocentric radius. However, the median mass of the parent clusters of our cluster-ERAGN is $10^{14.5}~\text{M}_\odot$, similar to that of the \citet{Moravec2020} sample. The redshift of two samples are also similar at z $\sim$ 1. These largely exclude mass and redshift differences between parent cluster samples as the reason for the lack of correlation. We do note that cluster centers and masses are calculated differently in these two surveys\footnote{In the MaDCoWS survey, the cluster center is defined as the peak flux in the infrared-selected galaxy overdensities, and the cluster mass is defined by the total number of infrared-selected galaxies and then calibrated to Sunyaev-Zel'dovich observations. }. The descriptions of cluster center and mass measures in the ORELSE survey are shown in Appendix \ref{app:eta}. 
On the other hand, the lack of correlation might partially be due to the difference in measuring size: in this work, we use radio maps with 5'' resolution, while \citet{Moravec2020} use radio images with 1-2'' resolution. The large beam size systematically enlarges the LAS of small sources. Although our larger synthesized beam systematically scales up the LAS of compact sources and introduces bias at the small-size end, measurements on the more extended sources are minimally affected, where the difference between high- and the low-density ERAGN manifests itself. 

Furthermore, we note that there are potential uncertainties in \citet{Moravec2020} study: the lack of redshift measurements of individual ERAGN which means that the true memberships of their ERAGN sample are highly uncertain, the existence of a large uncertainty in their cluster centers which will affect the R$_\text{proj}$ measurement, and the low completeness of their cluster catalog which can result in the incorrect assignment of an ERAGN to a parent cluster. 
Whereas, as is discussed in Appendix \ref{app:eta}, we ameliorate these uncertainties in our study. 
We know that even with high precision z$_\text{phot}$, a galaxy can shift in/out of the cluster/group when including an even small uncertainty of z$_\text{phot}$ (see Section \ref{sec:env} and Appendix \ref{app:eta}). This is illustrated, in the right panel of Figure \ref{sec:global}, where three ERAGN have their $\eta$ in the field region, but their errorbars extend to the intermediate cluster/group regions. In addition, 15 ERAGN in our sample are within 500 kpc ($\sim$1' at z = 1) relative to their parent structures in projected radius, but only 10 of those are actually located in the cluster/group environment as defined by $\eta \leq 2$.

As a result, instead of the clustocentric radius, the global environment $\eta$, defined in Section \ref{sec:ORELSE}, is a significantly better measurement of cluster/group environments as it eliminates projection effects \citep{Lucey1983, Postman1992}. 
As shown in the right panel of figure \ref{fig:global}, the LAS$_\text{norm}$ versus $\eta$, we observe a complete lack of large LAS systems (log(LAS$_{\text{norm}}$) $\leq$ 300 kpc) in the central regions, unlike other regions that always contain some. 
In addition, we find that ERAGN in the cluster/group centers are consistently more compact compared to those in the intermediate and field regions. 
The median of log(LAS$_\text{norm}$) for ERAGN in the cluster/group centers is 2.38 $\pm$ 0.02, while that for ERAGN in the intermediate and field regions are 2.48 $\pm$ 0.04 and 2.41 $\pm$ 0.03, respectively. 
The smallest error on the median of the log(LAS$_\text{norm}$) of central regions indicates that a smaller scatter in radio sizes in this region. 
As shown in Figure \ref{fig:global}, we include the uncertainty of $\eta$ for ERAGN-phot based on the uncertainty of z$_\text{phot}$ (see Appendix \ref{app:eta}). The uncertainty of z$_\text{phot}$ might shift three ERAGN from the field into the intermediate region. However there is no change between the intermediate and the central regions, because all but one source have spectroscopic redshifts. Thus, the comparisons presented in this section are not affected by our z$_\text{phot}$ uncertainties.

Again, we use Area$_\text{norm}$ to perform the same analyses and find the same results. No correlation is found between Area$_\text{norm}$ and R$_\text{proj}$ or $\eta$ for the high-density, cluster-hosted and full ERAGN samples. The median of log(Area$_\text{norm}$) for ERAGN in the cluster/group centers, intermediate and field regions are 4.32 $\pm$ 0.02, 4.35 $\pm$ 0.04 and 4.34 $\pm$ 0.05, respectively. The smallest scatter is shown in the cluster/group center regions. 

This series of comparisons have shown no correlations between the radio size and distance from the cluster center or global environment measurement in any sub-samples. However, we find that ERAGN in the cluster/group central regions are preferentially compact with the smallest scatter in size, compared to those in the cluster/group intermediate regions and the field. This result points to a plausible interpretation that the environment of cluster/group centers plays a significant role in confining the spatial extent of ERAGN, a conclusion consistent with simulations showing that relatively higher density ICM (e.g, that found in the center of a cluster) is more efficient in confining jets than lower density ICM (e.g. that found in the outskirts of a cluster; \citealp{Kaiser1997, Alexander2000, Kaiser2007}). 

\section{A Simple Numerical Simulation of Radio Jets in Different Environments}
\label{sec:simulation}

\begin{figure}
    \centering
    \includegraphics[width=\columnwidth]{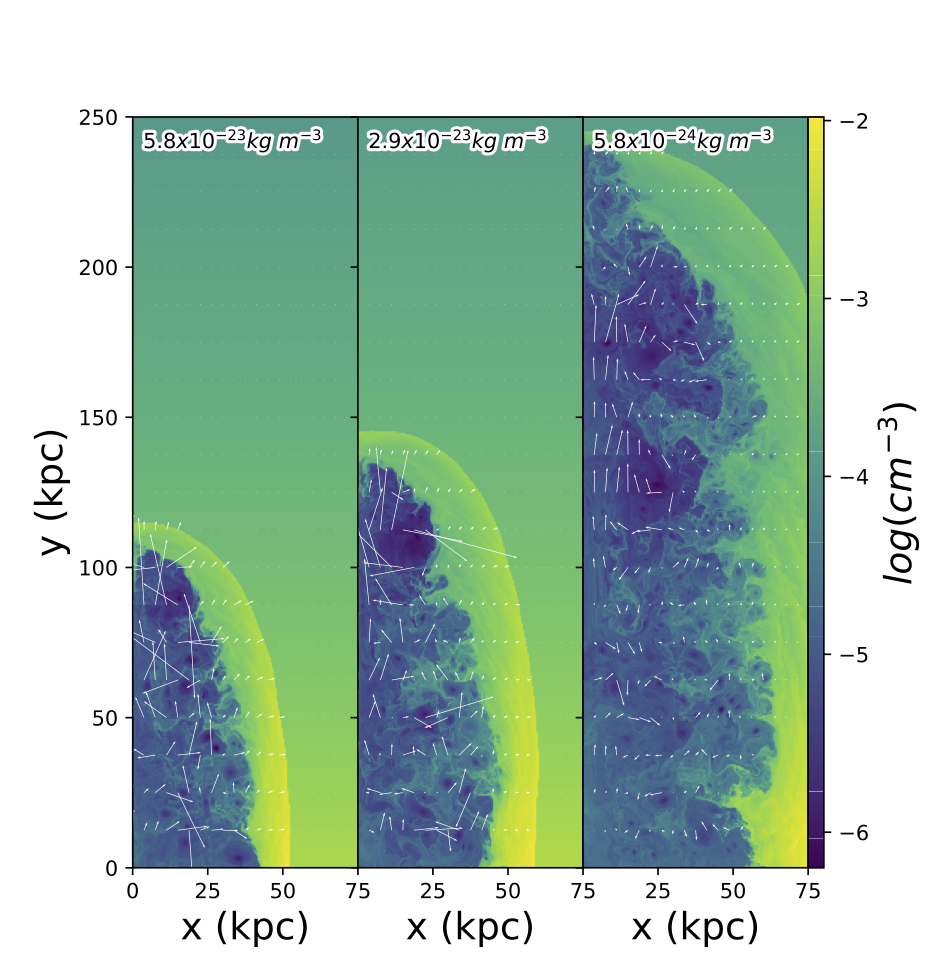}
    \caption{Jets simulation in three environments $5.8 \times 10^{-23}$, $2.9 \times 10^{-23}$, $5.8 \times 10^{-24} kg\ m^{-3}$. The colorful pattern indicates the number density of protons with unit of log($cm^{-3}$), and the white arrows indicate velocity.}
\label{fig:simulation}
\end{figure}

The environmental dependence of radio jets' properties has been investigated in environments from that of poor galaxy groups to rich clusters (e.g., \citealp{Krause2012, Hardcastle2013, Yates2018}). To summarize the results of these simulations, of which the majority concentrate on FR II sources, are: a) the gas pressure in a rich cluster is relatively high, which collimates the jet relatively early and thus produces a narrow beam; b) the high cluster density ensures that the lobes are bright radio emitters; c) in general, jets in groups are collimated at a later stage, subtend a wider angle and thus more difficult to observe. 

To delineate the size difference of jets solely due to variation in the ICM density, we conduct numerical simulations by propagating the same radio jet in environments with relatively low, intermediate and high densities, using the hydrodynamics (HD) code, \textit{PLUTO} \footnote{http://plutocode.ph.unito.it/} \citep{Mignone2007, Mignone2012}.
The setup of our simulation is similar to that in \citet{Yates2018}, with exception of jet's  kinetic power, initial opening angle and gas density.
We adopt a larger opening angle 30$^{\circ}$ to facilitate inclusion of FR I morphology \citep{Krause2012}, and a jet kinetic power of Q = 10$^{44.7}$ erg/s, which is converted from the median radio power in our sample (L$_\mathrm{{1.4GHz}}$ = 10$^{24.9}$ W Hz$^{-1}$) following the kinetic vs. radio power scaling relations given in \citet{Cavagnolo2010}. We choose three gas densities: $5.8 \times 10^{-23}$, $2.9 \times 10^{-23}$ and $5.8 \times 10^{-24}$ kg  m$^{-3}$, following the isothermal NFW gas density profile \citep{Navarro1996} at 0, 100 and 500 kpc from the center of a $3\times10^{14}$ M$_\odot$ cluster (\citealp{Yates2018} Figure 1). 
Figure~\ref{fig:simulation} shows the density plot of the radio jets in three density profiles when the jet switches off 60 Myr after launched. With the same evolving time elapsed, the effective area of the jet increases with decreasing density. As a consequence, the maximum length of the jet in the highest/intermediate density is $\sim$60\%/40\% less than that in the lowest density, supporting our observational conclusion that high ICM density may restrict the development of the length of jet. 

Admittedly, we only propagate a single burst of the jet; in the case of multiple bursts, the ICM gas may have been partially cleared by previous jets, and the jet we observe has actually propagated along a low density path. This scenario would weaken our interpretation. 
Note that the gas density changes along the y-axis only, so that we are not able to make any conclusion about the area of the jet nor make direct comparisons to real observations. 
Further complications exist due to the actual ICM density profile deviating from our isotropic simplification and to neglected factors (e.g. magnetic fields, density/temperature turbulence) that introduce uncertainties. Our simulations, presented here for heuristic purposes, focus solely on a qualitative reproduction of the dependence of radio sources’ jet spatial scale on the ICM density. Future work on more realistic simulations promises to provide more clues to how the observable properties of jets varies as a result of the jet-environment interaction.

\section{SUMMARY}
\label{sec:summary}

In this paper, by constructing a sample of 50 ERAGN at intermediate redshift (0.55 $\leq z \leq$1.3) from 12 fields in the ORELSE survey, we conduct a search for evidence of environmental effects on the ERAGN across a broad dynamical range of environments. All ERAGN have either spectropically-confirmed redshifts or accurate photometric redshifts, allowing us to quantify accurate local and global environmental measurements. Below is a summary of our fndings:

\begin{itemize}
	\item We find the radio size in high-density regions to be consistently more compact than that in low-density regions at a significance level of 4.5$\sigma$. 
	
	\item We find no significant correlations between the radio size and their host internal properties (i.e., stellar mass and color) and redshift, which eliminates the possibility of effects resulting from internal and evolutionary factors. 
	
	\item A positive correlation is found to exist between the size and the radio power of ERAGN, suggesting that radio power is a driver of the extent of the radio jet. The difference of radio size between high-density and low-density ERAGN persists when a radio power correction is applied.
	
	\item We find no correlations between the radio size and any environmental properties, i.e., local overdensity, distance from the cluster center or global environment measurement. 
	However, our analysis of the large-scale environment shows that ERAGN in the cluster/group central regions are preferentially compact with the smallest scatter in size, compared to those in the cluster/group intermediate regions and the field. 
	
	\item We perform a numerical simulation of radio jets in three different density environments, which shows the length of the jet decreases as the surrounding gas density increases, supporting our observational conclusion that the conditions at the centers of groups and clusters play an important role in confining the size of the radio sources in question. 

\end{itemize}

For future studies, we plan to follow up our ERAGN with higher resolution radio observations to better measure the size of radio jets. Future simulations that have more constraining power in 2-dimension and are more realistic will allow us to fully compare to our observational results. In addition, we plan to continue the investigation of ERAGN in clusters that have X-ray detections, to search for signs of the ICM variation which might allow us to prove or disprove this hypothesis. 

\section*{Acknowledgements}
We thank Paolo Padovani for helpful comments and suggestions. LS and GL acknowledge the grant from the National Key R\&D Program of China (2016YFA0400702), the National Natural Science Foundation of China (No. 11673020 and No. 11421303). WF acknowledge the NSFC Grants No. 11773024. 
This material is based upon work supported by the National Science Foundation under Grant No. 1411943. 
PFW acknowledges the support of an EACOA Fellowship from the East Asian Core Observatories Association. 
This study is based on data taken with the Karl G. Jansky Very Large Array which is operated by the National Radio Astronomy Observatory. The National Radio Astronomy Observatory is a facility of the National Science Foundation operated under cooperative agreement by Associated Universities, Inc. 
This work is based in part on observations made with the Spitzer Space Telescope, which is operated by the Jet Propulsion Laboratory, California Institute of Technology under a contract with NASA. 
SPIRE has been developed by a consortium of institutes led by Cardiff University (UK) and including Univ. Lethbridge (Canada); NAOC (China); CEA, LAM (France); IFSI, Univ. Padua (Italy); IAC (Spain); Stockholm Observatory (Sweden); Imperial College London, RAL, UCL-MSSL, UKATC, Univ. Sussex (UK); and Caltech, JPL, NHSC, Univ. Colorado (USA). This development has been supported by national funding agencies: CSA (Canada); NAOC (China); CEA, CNES, CNRS (France); ASI (Italy); MCINN (Spain); SNSB (Sweden); STFC, UKSA (UK); and NASA (USA). 
This work is based in part on data collected at the Subaru Telescope and obtained from the SMOKA, which is operated by the Astronomy Data centre, National Astronomical Observatory of Japan; observations made with the Spitzer Space Telescope, which is operated by the Jet Propulsion Laboratory, California Institute of Technology under a contract with NASA; and data collected at UKIRT which is supported by NASA and operated under an agreement among the University of Hawaii, the University of Arizona, and Lockheed Martin Advanced Technology centre; operations are enabled through the cooperation of the East Asian Observatory. When the data reported here were acquired, UKIRT was operated by the Joint Astronomy Centre on behalf of the Science and Technology Facilities Council of the U.K. 
This study is also based, in part, on observations obtained with WIRCam, a joint project of CFHT, Taiwan, Korea, Canada, France, and the Canada-France- Hawaii Telescope which is operated by the National Research Council (NRC) of Canada, the Institut National des Sciences de l'Univers of the Centre National de la Recherche Scientifique of France, and the University of Hawai'i. The scientific results reported in this article are based in part on observations made by the Chandra X-ray Observatory and data obtained from the Chandra Data Archive. 
The spectrographic data presented herein were obtained at the W.M. Keck Observatory, which is operated as a scientific partnership among the California Institute of Technology, the University of California, and the National Aeronautics and Space Administration. The Observatory was made possible by the generous financial support of the W.M. Keck Foundation. 
We wish to thank the indigenous Hawaiian community for allowing us to be guests on their sacred mountain, a privilege, without with, this work would not have been possible. We are most fortunate to be able to conduct observations from this site.



\appendix
\section{The Uncertainty of Global Environment Measurement} 
\label{app:eta}

In this paper, we adopt the global environment from \citet{Shen2017} as $$\eta = \mathrm{R_{proj}/R_{200}} \times |\Delta v|/\sigma_v,$$ where R$_\text{proj}$ is the distance of a given galaxy to the center of its parent structure, R$_{200}$ is the radius of its parent structure at which the matter density is 200 times the critical density, $\Delta v$ is the velocity offset of the galaxy from the systemic redshift of its parent structure, and $\sigma_v$ is the measured line-of-sight galaxy velocity dispersion of its parent structure. 
The parent structure is determined as the one with the smallest R$_\text{proj}$/R$_{200}$ in projected space and within $\pm$6000 km s$^{-1}$ in velocity space. If for a given galaxy, no cluster/group within $\pm$6000 km s$^{-1}$ are found, the parent structure is the one with the smallest $\eta$. See \citet{Shen2019} and \citet{Pelliccia2019} for more details on this calculation.
Here, we discuss all possible uncertainties on the global environment measurement. 

The final cluster/group catalog includes the preexisting spectroscopically confirmed clusters/groups \citep{Gal2008, Lemaux2019} and overdensity candidate regions \citep{Hung2020}. 
In the former case, the cluster/group centers are the i'th-luminosity-weighted centers of spectral member galaxies calculated using the method described in \citet{Ascaso2014}, while their systemic redshifts, velocity dispersions, and associated errors are computed using the method described in \citet{Lemaux2012}. 
In the latter case, their centers and systemic redshift are the barycenters detected in the Voronoi Monte-Carlo maps from \citet{Hung2020}, and the $\sigma_v$s are calculated from the estimated log(M$_\text{tot}$) according to equation 1 and 2 in \citet{Lemaux2012}, along with their associated errors. 
The cluster-hosted ERAGN (see discussion in Section \ref{sec:global}) reside in the massive end of the ORELSE structures with their median mass of $10^{14.5} \text{M}_\odot$ versus the median mass of $10^{13.9} \text{M}_\odot$ for all ORELSE structures.  
The purity and completeness of clusters/groups increase with increasing structure mass. 
For our new cluster/group overdensity candidates, the purity/completeness of our catalog are very high for M$_{tot} \geq 10^{13.5}$ M$_\odot$ (0.92/0.83 and 0.60/0.49 at z = 0.8 and z = 1.2, respectively) using mock catalogs with spectroscopic fractions of $\sim$20\%, a typical value for the ORELSE fields \citep{Hung2020}. It is therefore unlikely that we are assigning these ERAGN to the wrong structures because either they are not real structures or we have missed a true overdensity that is closer. 

The two methods described above to measure centers and systemic redshifts are well-correlated \citep{Hung2020}. In addition, both measurements of cluster centers are well-correlated to the X-ray centers for relaxed structures \citep{Rumbaugh2018}. 
As for the velocity dispersion, both methods have measured the error on $\sigma_v$ (see \citealp{Lemaux2012} and \citealp{Hung2020} for more details). 
We adopt a Monte-Carlo test to see if this uncertainty can affect on the determination of the parent structure and the calculation of the $\eta$ value. In each Monte-Carlo iteration, we assign a mock $\sigma_v$ to each cluster/group sampled from a Gaussian distribution centered on the measured $\sigma_v$ and its associated error. Then, we re-determine the parent structures and re-calculate the $\eta$ values for all ERAGN. In all 100 iterations, none of ERAGN change their parent structure. 
We define the uncertainty of log($\eta$) to be the average of the 16/84 percentile values of the log($\eta$) distributions. 
The uncertainty of log($\eta$) is very small with a median of 0.06 and a range of 0.04 $\sim$ 0.27. Thus, $\eta$ and R$_\text{proj}$ are both highly accurate. 

Furthermore, an accurate redshift is the key to confirming group/cluster membership and measuring precise environmental metrics (i.e., R$_\text{proj}$ and $\eta$).  
For galaxies that only have z$_\text{phot}$, we adopt a Monte-Carlo sampling method to eliminate the concern of their real membership due to the uncertainty of z$_\text{phot}$. 
For each galaxy, we sample 100 z$_\text{phot}$ from a Gaussian distribution centered on the peak with $\sigma$ derived from the the reconstructed probability density function of the photometric redshift estimated by the code Easy and Accurate Redshifts from Yale (\textsc{EAZY}: \citealp{Brammer2008}, also see \citealp{Tomczak2017} for more details on the z$_\text{phot}$ measurement). The median of $\sigma$ of z$_\text{phot}$ is 0.03 for ERAGN that only have z$_\text{phot}$, the same as the value of the full sample of galaxies with high-quality spectroscopic redshifts in ORELSE. This Monte-Carlo z$_\text{phot}$ sampling has also been used for the calculation of additional errors on radio power (Section \ref{sec:sample}), size (Section \ref{sec:LAS}), and local overdensity (Section \ref{sec:env}) for ERAGN-phot. 
Mock R$_\text{proj}$, $\Delta v$ and $\eta$ values are then recalculated with respect to their mock parent structures. 
As plotted in Figure \ref{fig:global}, the final R$_\text{proj}$, $\Delta v$ and $\eta$ are the median of the 100 mock values. The lower and upper limits of them are calculated as the 16/84 percentile values. The median of the uncertainty on log($\eta$) due to the uncertainty of z$_\text{phot}$ is 0.21, with a range of 0.01 $\sim$ 0.65. We note that varying the z$_\text{phot}$ does also change the parent structure for some ERAGN, which causes asymmetric distributions of mock R$_\text{proj}$ and $\eta$, manifested as asymmetric uncertainties in Figure \ref{fig:global}. 
We note that we run a parallel Monte-Carlo sampling allowing both $\sigma_v$ and z$_\text{phot}$ varying at the same time. The uncertainties of z$_\text{phot}$ contribute the most to the uncertainties of the R$_\text{proj}$ and $\eta$. Therefore, the uncertainties of global environment properties are calculated based solely on the uncertainty of z$_\text{phot}$ in this paper.

\section{The Catalog of ERAGN}
\label{sec:catalog}

\startlongtable
\centering
\begin{deluxetable}{lccccccccc}
\tablecaption{Properties of ERAGN \label{tab:catalog}} 
\tabletypesize{\footnotesize}
\tablehead{\colhead{Field} & \colhead{ID\tablenotemark{a}} & \colhead{Coords\tablenotemark{b}}  & \colhead{z\tablenotemark{c}} & \colhead{log(L$_\mathrm{1.4GHz}$)} & \colhead{LAS} & \colhead{log(Area)} &\colhead{log($1+\delta_\mathrm{gal}$)} & \colhead{log($\eta$)} & \colhead{local rms \tablenotemark{e}} \vspace{-3mm}\\ 
 & &  &  &  \colhead{log(W Hz$^{-1}$)} & \colhead{kpc} & \colhead{log(kpc$^2$)} & & & \colhead{$\mu$Jy} \vspace{-3mm}\\ 
 \colhead{(1)} & \colhead{(2)} & \colhead{(3)}  & \colhead{(4)} & \colhead{(5)} & \colhead{(6)} & \colhead{(7)} &\colhead{(8)} & \colhead{(9)} & \colhead{(10)}}
\startdata
SG0023&82&00h23m05.3s +04d21m25.1s&0.97*&25.16$\pm$0.02&131$\pm$20&4.01$\pm$0.01&0.49$^{+0.12}_{-0.07}$&1.97$^{+0.55}_{-0.30}$ & 12.8 \\
SG0023&127&00h23m13.8s +04d26m03.0s&0.72*&25.37$\pm$0.02&410$\pm$19&4.52$\pm$0.02&0.24$^{+0.05}_{-0.10}$&1.86$^{+0.26}_{-0.13}$ & 12.5 \\
SG0023&230&00h23m38.2s +04d17m00.4s&1.33&26.89$\pm$0.01&192$\pm$21&4.32$\pm$0.01&1.03 &2.62  & 10.5 \\
SG0023&256&00h23m44.9s +04d25m35.2s&0.75&24.72$\pm$0.01&104$\pm$18&3.77$\pm$0.01&0.23 &1.95 &10.7 \\ 
XLSS005&F4&02h25m54.9s -04d28m54.0s&1.07*&25.29$\pm$0.02&315$\pm$20&4.43$\pm$0.01&0.13$^{+0.14}_{-0.07}$&1.65$^{+0.29}_{-0.11}$ & 22.0 \\
XLSS005&F2&02h26m19.8s -04d25m31.4s&1.21*&25.97$\pm$0.04&199$\pm$21&4.20$\pm$0.01&0.29$^{+0.15}_{-0.24}$&1.17$^{+0.79}_{-0.54}$ &  13.1 \\
XLSS005&E47&02h27m02.2s -04d17m46.1s&1.19&24.72$\pm$0.02&316$\pm$21&3.53$\pm$0.01&0.72 & -0.81 & 7.0 \\
XLSS005&E49&02h27m33.5s -04d33m17.9s&0.62*&24.27$\pm$0.06&197$\pm$18&3.99$\pm$0.02&0.36$^{+0.08}_{-0.08}$&0.73$^{+0.67}_{-0.29}$ &15.0 \\
XLSS005&E46&02h27m43.2s -04d21m28.3s&0.73*&24.41$\pm$0.06&328$\pm$18&3.99$\pm$0.01&0.53$^{+0.14}_{-0.45}$&0.68$^{+0.55}_{-0.35}$ & 8.5 \\
RXJ0910&886&09h10m22.7s +54d22m49.4s&0.56*&24.92$\pm$0.03&194$\pm$51&4.17$\pm$0.02&0.25$^{+0.16}_{-0.02}$&1.44$^{+0.28}_{-0.09}$ &10.5 \\
RXJ1053&F10&10h52m25.6s +57d33m22.6s&0.62*&24.78$\pm$0.03&98$\pm$17&3.76$\pm$0.01&0.57$^{+0.03}_{-0.08}$ &2.08$^{+0.01}_{-0.01}$ & 6.7 \\
RXJ1053&1027&10h53m27.6s +57d45m43.9s&0.85*&25.15$\pm$0.02&203$\pm$19&4.17$\pm$0.01&0.22$^{+0.13}_{-0.12}$&2.27$^{+0.05}_{-0.05}$ & 6.4 \\
RXJ1053&E39&10h53m37.3s +57d42m40.6s&0.72*&24.43$\pm$0.04&272$\pm$18&3.56$\pm$0.01&0.40$^{+0.10}_{-0.19}$&2.26$^{+0.02}_{-0.02}$ & 5.5 \\
RXJ1053&1028&10h53m42.2s +57d44m36.7s&0.70*&24.89$\pm$0.02&210$\pm$18&4.23$\pm$0.01&0.31$^{+0.05}_{-0.13}$&2.38$^{+0.03}_{0.03}$ & 6.5 \\
Cl1137&785&11h37m18.7s +30d00m40.0s&0.62*&23.87$\pm$0.03&164$\pm$17&3.85$\pm$0.02&0.37$^{+0.05}_{-0.25}$&2.29$^{+0.02}_{-0.03}$ & 5.8 \\
Cl1137&783&11h37m33.0s +30d00m02.3s&0.96&25.06$\pm$0.01&117$\pm$20&3.85$\pm$0.01&2.03&-1.32 & 5.9 \\
Cl1137&784&11h37m33.8s +30d00m10.4s&0.99*&25.00$\pm$0.02&139$\pm$20&3.86$\pm$0.01&0.41$^{+0.15}_{-0.20}$&0.52$^{+0.36}_{-0.28}$ & 5.9 \\
SC1324&55&13h24m03.7s +30d43m40.4s&0.70&24.94$\pm$0.01&366$\pm$18&4.07$\pm$0.01&0.87&0.12 & 20.0 \\
SC1324&102&13h24m15.5s +30d46m44.9s&0.66*&24.32$\pm$0.02&146$\pm$18&3.47$\pm$0.01&0.35$^{+0.05}_{-0.01}$&1.37$^{+0.29}_{-0.83}$ & 16.0 \\
SC1324&115&13h24m20.1s +30d48m38.6s&1.20*&24.87$\pm$0.03&132$\pm$21&3.37$\pm$0.01&0.03$^{+0.02}_{-0.06}$&2.07$^{+0.46}_{-0.18}$ & 14.7 \\
SC1324&614&13h24m24.6s +30d21m21.4s&1.24*&26.10$\pm$0.02&196$\pm$21&3.79$\pm$0.01&0.04$^{+0.04}_{-0.03}$&2.49$^{+0.04}_{-0.05}$ & 12.2 \\
SC1324&148&13h24m31.1s +30d49m28.3s&0.58*&24.05$\pm$0.02&90$\pm$16&3.18$\pm$0.02&0.05$^{+0.01}_{-0.02}$&1.79$^{+0.09}_{-0.09}$ & 12.8 \\
SC1324&212&13h24m43.4s +30d49m38.0s&1.10&25.63$\pm$0.00&313$\pm$20&4.23$\pm$0.01&0.50&-0.81 & 13.3 \\
SC1324&699&13h24m44.9s +30d12m29.8s&1.08&25.19$\pm$0.01&221$\pm$20&3.81$\pm$0.01&0.24&2.19 &11.1 \\
SC1324&271&13h24m54.5s +30d49m15.9s&0.70&23.88$\pm$0.03&117$\pm$18&3.34$\pm$0.01&1.04&-0.45 & 13.1 \\
SC1324&323&13h25m06.2s +30d44m19.8s&0.68*&24.22$\pm$0.04&116$\pm$18&3.31$\pm$0.03&0.34$^{+0.06}_{-0.13}$&1.41$^{+0.44}_{-0.54}$ & 12.6 \\
SC1324&410&13h25m27.7s +30d42m52.4s&0.72*&24.22$\pm$0.03&115$\pm$18&3.44$\pm$0.02&-0.19$^{+0.12}_{-0.11}$ &1.69$^{+0.61}_{-0.40}$ &17.5 \\
Cl1350&617&13h50m50.1s +60d08m03.5s&0.81&24.41$\pm$0.03&98$\pm$19&3.53$\pm$0.01&0.88&-0.20 & 9.6 \\
Cl1350&327&13h50m59.6s +60d06m09.5s&0.80&25.43$\pm$0.01&156$\pm$19&3.98$\pm$0.01&1.51&-1.01 & 9.6 \\
Cl1429&F3&14h28m18.7s +42d50m35.2s&0.60*&24.15$\pm$0.09&61$\pm$17&3.29$\pm$0.03&-0.70$^{+0.09}_{-0.08}$&1.98$^{+0.19}_{-0.13}$ & 22.0 \\
Cl1429&F10&14h28m21.5s +42d35m06.4s&1.12*&25.70$\pm$0.04&118$\pm$21&3.77$\pm$0.01&-0.60$^{+0.14}_{-0.08}$&1.87$^{+0.38}_{-0.18}$ & 20.0 \\
Cl1429&E25&14h28m38.0s +42d44m39.9s&0.62*&25.25$\pm$0.05&196$\pm$18&4.08$\pm$0.01&0.13$^{0.04}_{-0.04}$ &1.78$^{+0.23}_{-0.15}$ &18.0 \\
Cl1429&F9&14h28m46.2s +42d38m13.7s&0.83&25.05$\pm$0.02&85$\pm$19&3.56$\pm$0.01&0.51&0.46 & 16.0 \\
SC1604&409&16h04m04.2s +43d23m57.7s&0.68*&24.64$\pm$0.02&87$\pm$18&3.67$\pm$0.01&0.46$^{+0.07}_{-0.32}$ &1.56$^{+0.38}_{-0.19}$ & 8.2 \\
SC1604&427&16h04m06.4s +43d18m08.0s&0.92&24.29$\pm$0.01&99$\pm$20&3.65$\pm$0.01&1.07&0.87 & 6.9 \\
SC1604&F11&16h04m23.6s +43d14m07.1s&0.86&24.98$\pm$0.03&131$\pm$19&3.89$\pm$0.01&1.42&-1.50 & 7.0 \\
SC1604&557&16h04m24.8s +43d04m25.4s&0.90&23.95$\pm$0.04&96$\pm$19&3.63$\pm$0.01&1.45&-0.78 & 8.1 \\
SC1604&561&16h04m25.1s +43d04m50.5s&0.90&24.34$\pm$0.01&122$\pm$19&3.82$\pm$0.01&2.05&-1.61 & 8.1 \\
SC1604&574&16h04m26.8s +43d15m04.1s&0.86&24.18$\pm$0.01&81$\pm$19&3.55$\pm$0.01&1.20&-0.54 & 7.0 \\
SC1604&620&16h04m30.8s +43d16m36.2s&0.90&24.32$\pm$0.02&153$\pm$19&3.76$\pm$0.01&0.34&0.69 & 7.1 \\
SC1604&697&16h04m39.9s +43d21m14.2s&0.92&25.10$\pm$0.00&110$\pm$20&3.79$\pm$0.01&1.19&-1.19 & 7.7 \\
SC1604&701&16h04m40.4s +43d22m20.4s&1.27&25.13$\pm$0.00&99$\pm$21&3.79$\pm$0.01&0.63&-0.27 & 7.8 \\
RXJ1716&E18&17h15m03.9s +67d12m03.2s&0.70*&24.87$\pm$0.07&271$\pm$24&4.12$\pm$0.04&-0.19$^{+0.03}_{-0.10}$ &1.71$^{+0.55}_{-0.20}$ & 16.0 \\
RXJ1716&E17&17h16m37.0s +67d08m29.4s&0.790&26.82$\pm$0.04&336$\pm$19&4.50$\pm$0.01&1.14&-0.10 & 16.1 \\
RXJ1716&567&17h16m56.9s +67d09m04.0s&0.81*&24.21$\pm$0.05&105$\pm$19&3.46$\pm$0.02&1.24$^{+0.37}_{-0.31}$ &0.03$^{+0.48}_{-0.20}$ & 14.8 \\
RXJ1757&152&17h56m42.3s +66d33m34.0s&1.34*&26.63$\pm$0.13&221$\pm$22&3.98$\pm$0.03&0.10$^{+0.08}_{-0.08}$ &2.16$^{+0.44}_{-0.16}$ & 11.7 \\
RXJ1821&196&18h21m07.2s +68d26m09.7s&1.30*&24.88$\pm$0.03&132$\pm$21&3.47$\pm$0.01&0.08$^{+0.04}_{-0.06}$ &1.53$^{+0.14}_{-0.07}$ & 9.4 \\ 
RXJ1821&255&18h21m30.7s +68d29m28.0s&0.82&24.60$\pm$0.02&151$\pm$19&3.82$\pm$0.01&1.40&-0.65 & 9.9 \\
RXJ1821&269&18h21m37.1s +68d27m50.7s&0.82&24.41$\pm$0.01&111$\pm$19&3.38$\pm$0.01&1.91&-1.49 & 9.4 \\
RXJ1821&294&18h21m47.4s +68d21m03.2s&1.28&24.71$\pm$0.01&126$\pm$21&3.32$\pm$0.01&-0.02&2.08 & 9.3 \\
\enddata
\tablenotetext{}{(1) (2) the field and ID of ERAGN. ERAGN detected by eye start with an “E”; (3) the R.A. and Decl. of the radio centers of ERAGN; (4) the redshift of ERAGN with z$_\mathrm{phot}$ marked by *. (5) the radio power of radio sources, see Section \ref{sec:sample}; (6) (7) the extent of radio sources and their associated error, see Section \ref{sec:LAS}; (8) (9) the local and global environment measurements, described in Section \ref{sec:env} with more details in Appendix \ref{app:eta} for their error calculation; (10) the local rms is measured in a 2' $\times$ 2' empty region close to the radio source in the radio image. The size of the region is chosen to eliminate the local variations and not to include other radio sources. }
\end{deluxetable} 



\end{document}